\input psfig
\documentstyle[twoside,fleqn,espcrc2]{article}

\newcommand{\AmS}{{\protect\the\textfont2
  A\kern-.1667em\lower.5ex\hbox{M}\kern-.125emS}}

\title{ Truncated Determinant  Approach to Light Dynamical Quarks}
\author{ A.~Duncan\address{Dept. of Physics and Astronomy,University of Pittsburgh,
 Pittsburgh, PA 15260},%
  E.~Eichten\address{ Fermilab, PO Box 500, Batavia, IL60510},%
  and 
  H.~Thacker\address{Dept. of Physics, University of Virginia, 
 Charlottesville, VA 22901}}
\begin{document}
\begin{abstract}
  Virtual quark effects in lattice QCD are studied in an approach where the infrared
 modes are included in a precise and gauge-invariant manner. At fixed physical
 volume the required number of modes does not increase in the continuum limit, and
  the acceptance is tolerable even for very light
 quark masses.  Topological charge distributions at varying quark mass,
 and the static quark
 potential, showing the onset of string-breaking, are presented in both QED2 and QCD4.
 
\end{abstract}

\maketitle

\section{Role of Infrared Dirac Modes in Nonperturbative Lattice QCD}

  It is well known that the quenched approximation seriously mutilates the infrared Dirac
 structure of lattice QCD. Among the many undesirable effects of this truncation of the
 theory one finds\\
(1) Nonintegrable quark propagator singulariities render the Wilson-Dirac quenched
 path integral {\em undefined} \cite{qnchart}, a pathology which surfaces in explicit
 Monte Carlo simulations in the guise of ``exceptional configurations" leading to uncontrolled
 statistical fluctuations.\\
(2) Quenched chiral logarithms complicate the extraction of reliable chiral extrapolations.\\
(3) Nontrivial topology (i.e. nonzero topological charge) is not suppressed in the light
 quark limit.\\
(4) Characteristic physical phenomena induced by virtual quark pairs- string-breaking, 
 rho decay, etc- are absent.\\
   Inclusion of the infrared quark modes, characterized as the low eigenvalues
 of the hermitian operator $H\equiv\gamma_{5}(D\!\!\!/-m)$, allows a precise gauge invariant truncation \cite{Lat97}  which interpolates in a natural way between the quenched and fully
 unquenched theory, eliminating all  the aforesaid problems. The spectrum of  $H$ can be thought of as characterizing the gauge-invariant
 off-shellness of the quark field, as the eigenvalues  are  gauge-invariant and
 reduce in the free theory to $\pm\sqrt{p^{2}+m^{2}}$ for a quark mode of Euclidean momentum $p$.
 We shall discuss simulations in which all modes up to a cutoff somewhat greater than (typically,
 about twice) $\Lambda_{QCD}$ are
 included in the quark determinant part of the full Boltzmann weight. It turns out that accurate
 inclusion of  the remaining ultraviolet modes \cite{esti} requires a cutoff  at
  a fixed number of eigenvalues rather than  in energy.

\section{Implementing a Gauge Invariant Truncation of the Quark Determinant}

  The spectrum of the hermitian Dirac operator $H$ consists of eigenvalues
 $\lambda_{i}(A)$ which are  gauge-invariant functions of the gauge field $A$.
 For QCD4 the dimension of $H$ is ${\cal N}=12V$, with $V$ the lattice volume,
 and it is convenient to order the spectrum with the index $i$ running from $-{\cal N}/2$ to
 ${\cal N}/2$ with $\lambda_{i}<\lambda_{i+1}$.
 With the conventional lattice normalization the spectrum extends from $-(1+8\kappa)$ to
 $+(1+8\kappa)$ (roughly, from -2 to +2) with the physical branch extending up to 
 $|\lambda|\simeq 0.5$. We shall describe the results of simulations in which the full
 quark determinant contribution is replaced by ${\cal D}_{N_{\lambda}}(A)\equiv\prod_{i=-N_{\lambda}}
^{+N_{\lambda}}\lambda_{i}(A)$.  For QCD4 we choose $N_{\lambda}\simeq 50$,
 corresponding to $\lambda_{N_{\lambda}}\simeq$ 370 Mev, in other words, somewhat larger
 than the low energy scale $\Lambda_{QCD}$. Low eigenvalues are obtained by
 the following Lanczos scheme:\\
(1) Starting with a randomly chosen initial vector $v_1$, an orthonormal sequence $v_1,v_2,v_3,..v_{N_L}$ is generated by the standard recursion:\\
$\;\;\;\;v_{k+1}=\frac{1}{\beta_{k}}Hv_k-\frac{\beta_{k-1}}{\beta_{k}}v_{k-1}-\frac{\alpha_{k}}{\beta_{k}}v_{k}$\\
with the constants $\alpha_{k},\beta_{k}$ determined from overlaps of generated vectors. In the
 basis of the $v_{i}$, $H$ is tridiagonal. The corresponding real symmetric
 tridiagonal matrix $T_{N_L}$ has the $\alpha_{k}$ on the diagonal and the $\beta_{k}$ on the 
 sub (and super) diagonal. \\
(2) A Cullum-Willoughby sieve \cite{Cullum} is applied to remove spurious eigenvalues.\\
(3)  The remaining ``good" eigenvalues converge most rapidly in the needed infrared portion
 of the spectrum. The stability and accuracy of the converged
 eigenvalues has been checked extensively by gauge transforming the gauge field. \\
(4) The diagonalization of the $T_{N_L}$ matrix (typically, of order 10,000 in the QCD case)
 can be completely parallelized using the Sturm sequence property \cite{golub}
 of tridiagonal matrices
 which allows nonoverlapping parts of the spectrum to be independently extracted by a bisection
 procedure.

\begin{figure}
\psfig{figure=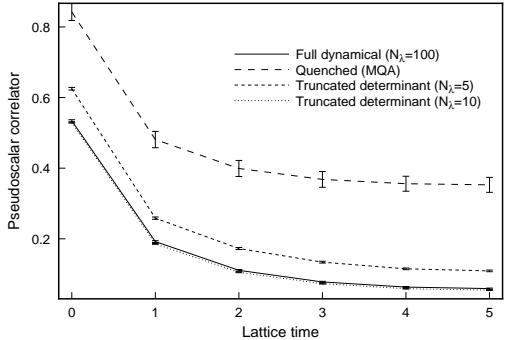,
width=0.95\hsize}
\vspace{-0.35in}
\caption{Pseudoscalar correlators in QED2}
\label{fig:pionqed}
\end{figure}
 
\section{Truncated Determinant Simulations in QED2}

  We have tested the truncated determinant approach in two-dimensional abelian
 gauge theory (QED2) on 10x10 and 16x16 lattices at $\beta=$4.5 and for bare
 quark masses $m_0=$0.06 and 0.10.  Superrenormalizability
 implies that higher momentum modes are essentially inert (the product of the 
 upper 90\% of the spectrum is practically constant) and one finds
 complete agreement between exact dynamical simulations and those using only
 the $2N_{\lambda}=$20 lowest eigenmodes of the (hermitian) Wilson-Dirac operator. 
\begin{figure}
\psfig{figure=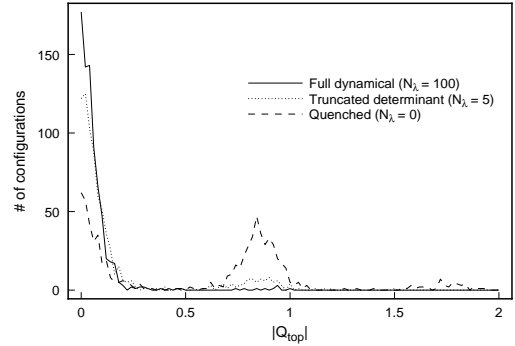,
width=0.95\hsize}
\vspace{-0.43in}
\caption{$Q_{top}$ histogram in QED2}
\label{fig:topqed}
\end{figure}
The simulation
 procedure used was as follows. A pure gauge update consisting of 5 metropolis
 sweeps was followed by an accept-reject step using the recomputed value of the 
 truncated determinant $ {\cal D}_{N_{\lambda}}(A)$. Strictly speaking \cite{Toussaint}
 detailed balance only holds for the pure gauge step if  links or sets of noninterfering
 links are updated in random order. For our QED2 runs we have found no statistically 
 significant differences between the results obtained with ordered or random link
 updates. (A fully parallel code incorporating detailed balance 
  will be used in all future truncated determinant simulations). The acceptance ratio for this
 procedure is tolerable (typically from 20-40\%) because the typical fluctuation in  
 $\log{ {\cal D}_{N_{\lambda}}(A)}$ after a  metropolis sweep is of order unity.
 (This is also the case in QCD4 with $N_{\lambda}$  chosen large enough to 
 include all the low energy chiral physics).
\begin{figure}
\psfig{figure=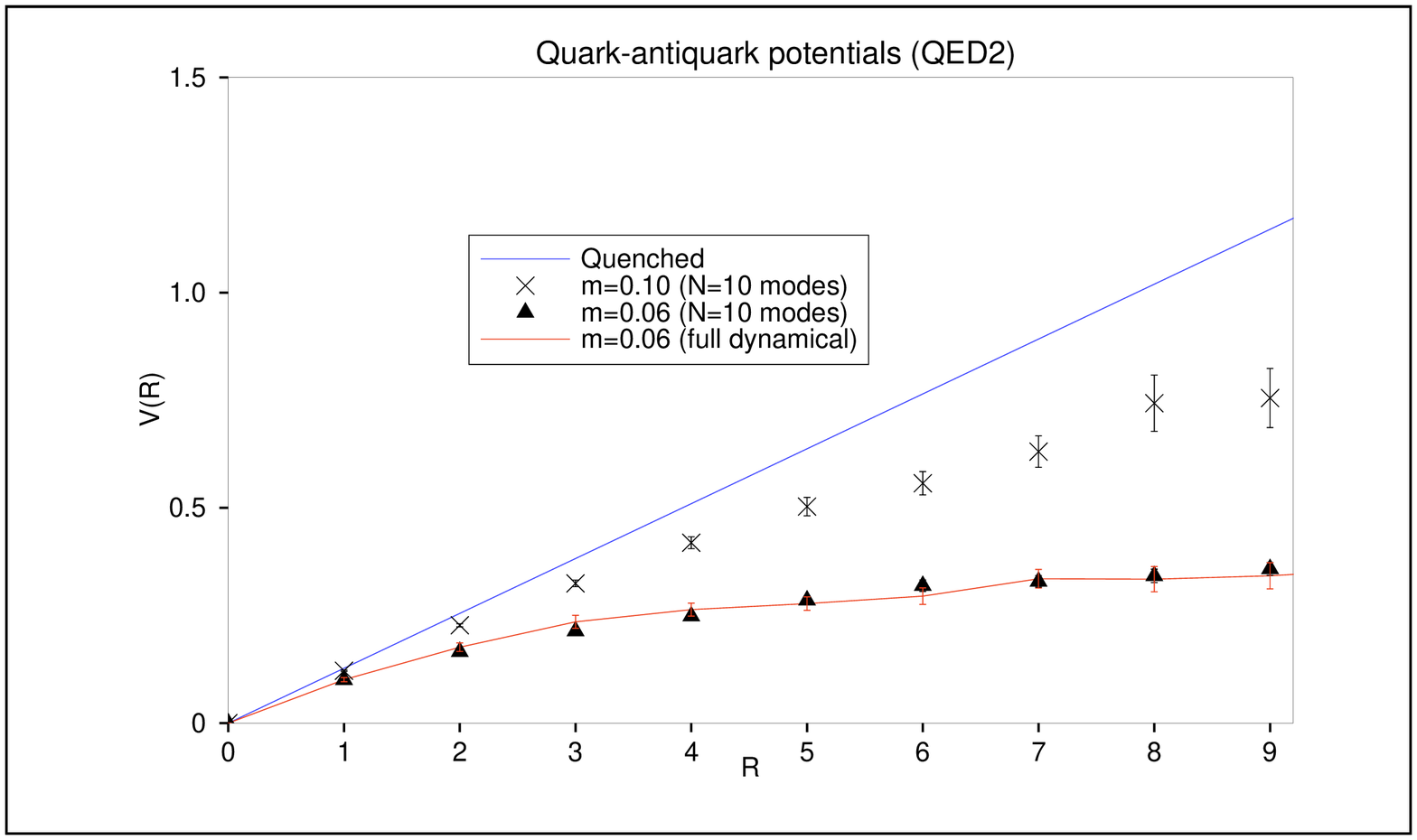,
width=0.95\hsize}
\vspace{-0.43in}
\caption{Quark-antiquark potentials in QED2}
\label{fig:qedstring}
\end{figure}

  Some results obtained from truncated determinant simulations in QED2 are shown in Figs. 1-3. 
 In Fig(1) the ``pion" (i.e. pseudoscalar) correlators are shown for a very light quark: inclusion of the lowest 20 eigenvalues leads to
 correlators essentially identical to those in the fully unquenched theory. 
In Fig(2) the
 topological charge ($Q_{top}$) distribution is shown for the same parameters in quenched, 
 unquenched and truncated determinant versions of the theory.  The suppression of
 nonzero topological charge is basically complete when only 5\% of the spectrum is 
 included in the determinant. Fig.3 shows the onset of string breaking in the  quark-antiquark
 potential measured from Wilson loops on a 16x16 lattice. 
\begin{figure}
\psfig{figure=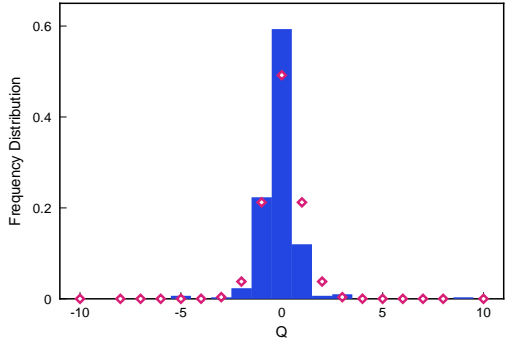,
width=0.95\hsize}
\vspace{-0.43in}
\caption{$Q_{top}$ histogram in QCD4 for  $\kappa$=0.1597 (diamonds- chiral result)}
\label{fig:topqcd}
\end{figure}
\section{Truncated Determinant Simulations in QCD4}

  We have  used the truncated determinant approach in QCD4
 on a 12$^3$x24 lattice at $\beta$=5.9 for three kappa values, $\kappa$=0.1570, 0.1587 and
 0.1597. Pion correlators were measured in order to perform a chiral extrapolation and
 determine the critical $\kappa_c$.  100 modes (up to $\simeq$370 MeV)  were included
 in the quark determinant, with either one or two complete
 Metropolis pure gauge sweeps between determinant accept-reject steps (with acceptance
 varying betwen 20\% and 50\%). The topological charge can be extracted from an anomalous
 chiral identity, and is related to the trace of the inverse of  $H=\gamma_5(D\!\!\!/-m)$. This  spectral sum converges rapidly:
 summing the lowest 100 eigenvalues  measures the topological
 charge (as per this definition) to a few percent. The histogram of measured topological
 charge in the lightest case, $\kappa$=0.1597 (corresponding to a pion mass of about 280 MeV)
 is shown in Fig.4, together with the analytic prediction for this distribution following from a chiral analysis \cite{Leut}.
\begin{figure}
\psfig{figure=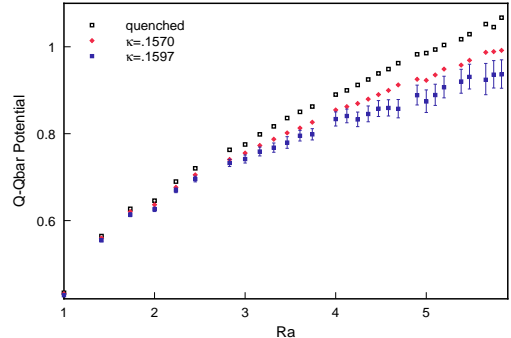,
width=0.95\hsize}
\vspace{-0.43in}
\caption{Static potential for quenched and dynamical QCD4. Error bars are largest for
 lightest mass and only shown for this case}
\label{fig:qqbar}
\end{figure}

In Fig.5 the static quark-antiquark potential (obtained
 from the correlation of Wilson lines in Coulomb gauge) shows a clear screening effect although true asymptotic flattening presumably
 occurs at distances where statistical fluctuations and finite volume effects dominate.

\vspace{-0.1in}



\begin{thebibliography}{99}
\bibitem{qnchart}  W. Bardeen, A. Duncan, E. Eichten and H. Thacker, hep-lat/9806002.
\bibitem{Lat97}  Talk of A. Duncan, Lattice 97, Proc. Suppl. Nucl. Phys. B63, 811 (1998);
 A. Duncan, E. Eichten and H. Thacker, hep-lat 9806020.
\bibitem{esti} Talk of E. Eichten, this conference.
\bibitem{Cullum}  J. Cullum and R.A. Willoughby, J. Comp. Phys. 44, 329 (1981).
\bibitem{golub}  G.H. Golub and C.F. Loan, {\em Matrix Computations}, 2nd edition 
 (Johns Hopkins, 1990).
\bibitem{Toussaint}  We thank D. Toussaint for explaining the connection of the detailed 
 balance condition to the link update procedure.
\bibitem{Leut} H. Leutwyler and A. Smilga, Phys. Rev. D46, 5607 (1992).
\end{thebibliography}
\end{document}